\documentclass[10pt]{iopart}

\usepackage{iopams}
\usepackage{graphicx}
\begin{document}


\title{High field magnetisation of the frustrated one dimensional quantum antiferromagnet LiCuVO$_{4}$}

\author{M. G. Banks$^1$, F. Heidrich-Meisner$^2$,  A.
Honecker$^3$, H. Rakoto$^4$, J.-M. Broto$^4$ and R. K. Kremer$^1$}

\address{$^1$ Max-Planck-Institute f\"{u}r Festk\"{o}rperforschung,
Heisenbergstrasse 1, 70569 Stuttgart, Germany}

\address{$^2$ Materials Science and Technology Division,
Oak Ridge National Laboratory, Oak Ridge, 37831 Tennessee, USA and
Department of Physics and Astronomy, University of Tennessee,
Knoxville, Tennessee 37996, USA}

\address{$^3$ Institut f\"{u}r Theoretische Physik, Universit\"{a}t G\"ottingen,
Friedrich-Hund-Platz 1, 37077 G\"{o}ttingen, Germany}

\address{$^4$ Laboratoire National des Champs Magn\'{e}tiques
Puls\'{e}s, 31432 Toulouse, France}

\ead{m.banks@fkf.mpg.de}

\begin{abstract}
We have investigated the high field magnetisation of the
frustrated one dimensional compound LiCuVO$_{4}$. In zero field,
LiCuVO${_4}$ undergoes long range antiferromagnetic order at
$T{_N}$ $\approx$ 2.5 K with a broad short range Schottky type
anomaly due to one dimensional correlations in the specific heat
at 32 K. Application of a magnetic field induces a rich phase
diagram. An anomaly in the derivative of the magnetisation with
respect to the applied magnetic field is seen at $\sim$ 7.5 T with
$H \parallel c$ in the long range order phase. We investigated
this in terms of a first experimental evidence of a middle field
cusp singularity (MFCS). Our numerical DMRG results show that in
the parameter range of LiCuVO${_4}$ as deduced by inelastic
neutron scattering (INS), there exists no MFCS. The anomaly in the
derivative of the magnetisation at $\sim$ 7.5 T is therfore
assigned to a change in the spin structure from the $ab$ plane
helix seen in zero field neutron diffraction.

\end{abstract}


\section{Introduction}

In the course of the vivid search for a theoretical understanding
of high-$T_c$ oxocuprate superconductors, the magnetic properties
of low-dimensional quantum $S$=$\frac{1}{2}$ antiferromagnetic
(afm) systems play a prominent role. Attention has been focussed
on theoretical and experimental investigations especially of quasi
one-dimensional afm systems since it became clear that electronic
phase separation creating e.g.  doping induced `stripe-like'
aggregates may be essential in the formation of the
superconducting phase
\cite{rk:kremer,rk:kremer2,jt:stripes,jz:domain}. A larger number
of new quasi one-dimensional copper oxides structurally closely
related to high-$T_c$ oxocuprates have since been prepared and
their magnetic properties were investigated in detail
\cite{dj:cuprate}. Unusual ground-state properties have been seen
to evolve due to the proximity of such systems to quantum
criticality via e.g. a considerable sensitivity to higher-order
effects in the exchange coupling but also to coupling to lattice
or charge degrees of freedom.

Most of the quasi one-dimensional copper oxide systems
investigated so far contain more or less \textit{stretched}
Cu-O-Cu bonds with bonding angles close to $\sim$180$^\circ$ which
leads to superexchange with exchange constants of the order of 100
meV \cite{dj:twoleg}, very similar to that found in the undoped
parent compounds of the high-$T_c$ oxocuprate superconductors
\cite{rc:la2cuo4}.

\begin{figure}
\begin{center}
\includegraphics[width=8cm]{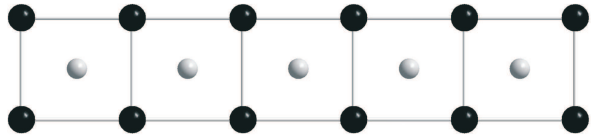}
\caption{Schematic view of a CuO$_2$ ribbon chain made up of
edgesharing CuO$_4$ squares. The grey spheres represent the
Cu$^{2+}$ ions, the black spheres the O$^{2-}$ ions.}
\label{fig:ribbon}
\end{center}
\end{figure}

Less broadly investigated are quasi one-dimensional systems which
contain isolated CuO$_2$ ribbon chains made up of edge-sharing
(slightly stretched or deformed) CuO$_4$ squares (see Fig.
\ref{fig:ribbon}). For an isolated CuO$_2$ ribbon chain, the spin
exchange interactions of interest are the nearest-neighbor (NN)
interaction $J_1$, which takes place through the two Cu-O-Cu
superexchange paths, and the next-nearest neighbor (NNN)
interaction $J_2$, which takes place through the two Cu-O $\cdot
\cdot \cdot$ O-Cu super-superexchange (SSE) paths
\cite{ym:cuo,mw:licuvo4}.

A broadly investigated system which contains such CuO$_2$ ribbon
chains which are connected via GeO$_4$ tetrahedra is e. g. the
inorganic spin Peierls compound CuGeO$_3$ ($T_{SP} \approx$ 14 K)
\cite{vo:voll,mh:hase}. The importance of the NNN interaction in
CuGeO$_3$ has first been discussed by Castilla \textit{et al.}
\cite{gc:cugeo3}.

Quasi one-dimensional systems containing isolated CuO$_2$ ribbon
chains offer the possibility to study one-dimensional compounds
with frustrated magnetism. Magnetic frustration is brought about
by the competition of the NN and NNN interaction. If $J_1$ and
$J_2$ are both afm, frustration primarily does not emerge from the
geometry of the lattice as for example is realized with the
triangular or the Kagom\'{e} in 2D, or with a pyrochlore lattice
in 3D, but rather from the topology of the Hamiltonian which can
be mapped to that of a zigzag chain with NN interactions only.

It can easily be shown that one-dimensional classical magnets with
competing interactions may result in a helicoidal ground state. In
the quantum case with $S = \frac{1}{2}$, strong frustration can
not only lead to a helicoidal ground state but also to gapped spin
liquid phases or states with local spin correlations, depending on
the frustration ratio $\alpha$ (= $J_{2}$/J$_{1}$)
\cite{sw:white,B:groundstate}.

Theoretically, much work has been carried out on a $J_{1}$-$J_{2}$
model for a one-dimensional Heisenberg $S = \frac{1}{2}$ afm ever
since the discovery of non-classical ground states
\cite{B:groundstate,mg:nnn}. Calculation of the thermodynamic
properties resulted in possible novel excitations of a domain wall
type \cite{bs:domain}. More recently, much work has been carried
out employing the density-matrix renormalization group (DMRG)
\cite{sw:dmrg,km:dmrg} which does not suffer from the sign problem
as in the case of Monte Carlo (MC) simulations, and also gives
results on substantially larger system sizes than accessible with
full diagonalization techniques.


Special attention has been paid to the magnetisation process of a
one-dimensional chain with both afm ($J_{1}$, $J_{2} >$ 0)
competing NN and NNN exchange interactions. For special values of
$\alpha$, so-called additional `middle-field cusp singularities'
(MFCS) at magnetic fields significantly below the saturation field
can appear in the magnetisation ($M-H$ curve). The origin of the
MFCS is a double-minimum shape of the energy dispersion of the
low-lying excitations.

The $M-H$ curve of an antiferromagnetic zigzag chain at zero and
finite temperatures was calculated using DMRG in the thermodynamic
limit for various values of $\alpha$ by Okunishi and collaborators
\cite{kh:nnnmag,ok:mag,nm:nnnmag2,ok:cusp}. At zero temperature a
MFCS was seen in the $M-H$ curve for $\alpha >$ 0.25. For $\alpha
\leq$ 0.25, the dispersion of a one down spin has one minimum. For
$\alpha > $ 0.25, we have at $k$ = $\pi$, a local maximum and two
minima that appear at either side of the maximum.

Recently, it was shown that the compound LiCuVO$_{4}$ represents
an example of a quasi one-dimensional Heisenberg $S = \frac{1}{2}$
afm in which such a frustrated situation is realized. LiCuVO$_4$
($\equiv$V[LiCu]O4) in the standard spinel notation) crystallizes
in an orthorhombically distorted inverse spinel structure, with
the non-magnetic V$^{5+}$ ions at the tetrahedrally coordinated
sites and Li$^+$ and Cu$^{2+}$ (3$d^9$ configuration) occupying in
an ordered way the octahedrally coordinated sites
\cite{ad:durif,ml:lafon}. The Jahn-Teller distorted CuO$_6$
octahedra connect via trans edges to form infinite Cu$^{2+}$
chains along the crystallographic $b$ direction leaving two nearly
rectangular ($\sim$95$^{\circ}$) Cu-O-Cu super-exchange paths
between NN Cu ions. The resulting CuO$_2$ ribbons are connected by
VO$_4$ tetrahedra that alternate up and down along the chain
direction. LiCuVO$_4$ exhibits the typical features of a quasi
one-dimensional Heisenberg $S = \frac{1}{2}$ afm, e. g. a broad
short range Schottky type anomaly in the specific heat at 32 K
\cite{mb:cplicuvo4} and a broad short range ordering maximum in
the magnetic susceptibility. In fact, initially LiCuVO$_4$ was
described as a quasi one-dimensional Heisenberg $S = \frac{1}{2}$
afm with NN neighbor interactions only
\cite{gb:blass,my:licuvo4cp,av:licuvo4cp,ck:licuvo4,tt:licuvo4,krug:licuvo4}.

LiCuVO$_{4}$ undergoes long range afm order at $\sim$2.5 K due to
interchain interactions.  The magnetic structure of LiCuVO$_{4}$
was first determined by single crystal neutron diffraction and
found to realize an incommensurate helix polarized along the chain
direction \cite{BG:maglicuvo4}. The incommensurability was
proposed to be caused by a scenario of frustration involving a NN
and NNN interaction along the chain. Subsequent inelastic neutron
scattering determined the NN exchange to be ferromagnetic ($J_{1}$
$\approx$ -12 K), rather than the expected afm interaction, and
the NNN exchange being substantially larger and antiferromagnetic
($J_{2}$ $\approx$ 41 K), thus confirming unquestionably this
scenario of magnetic frustration along the chain
\cite{ME:exlicuvo4}.

Here we report high field magnetisation measurements on single
crystals of LiCuVO$_{4}$ in the magnetically ordered phase. We
found an anomaly in the derivative of the magnetisation at $\sim$
7.5 T with $H \mid \mid c$. Using the exchange integrals from INS
experiments we have calculated the magnetisation process for
several system sizes by means of DMRG calculations. Our results
suggest that the anomaly seen at 7.5 T in the derivative of the
magnetisation is most likely \textit{not} due to a MFCS but rather
originates from a change of the magnetic structure. The true
nature of the phase transition and the magnetic structure in this
high field phase is unclear at present.

\section{Experimental}

Single crystals of LiCuVO$_4$ (space group Imma) were grown from
solutions of LiCuVO$_4$ in a LiVO$_3$ according to the procedures
described elsewhere \cite{pa:licuvo4growth}. Magnetisation was
measured down to 1.8 K in a magnetic field up to 7 T. Heat
capacity and susceptibility measurements both revealed a
transition at 2.5 K due to antiferromagnetic long range order.
High field magnetisation was carried out until a maximum field of
55 T, at the Laboratoire National des Champs Magn\'{e}tiques
Puls\'{e}s Toulouse, France.

\section{Results and discussion}

\begin{figure}
\begin{center}
\includegraphics[bb = 16 13 310 230]{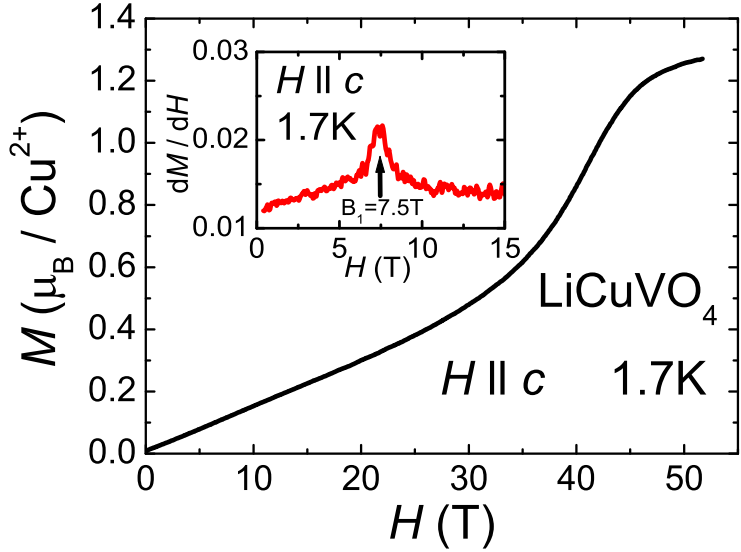}
\caption{High field magnetisation of LiCuVO$_{4}$ with $H \mid
\mid c$ at 1.7 K. Insert: derivative of the magnetisation with
respect to the applied field for $H \mid \mid c$ showing a clear
anomaly at $\sim$ 7.5 T.} \label{fig:licuvo4hfc}
\end{center}
\end{figure}

\begin{figure}
\begin{center}
\includegraphics[bb = 16 13 314 241]{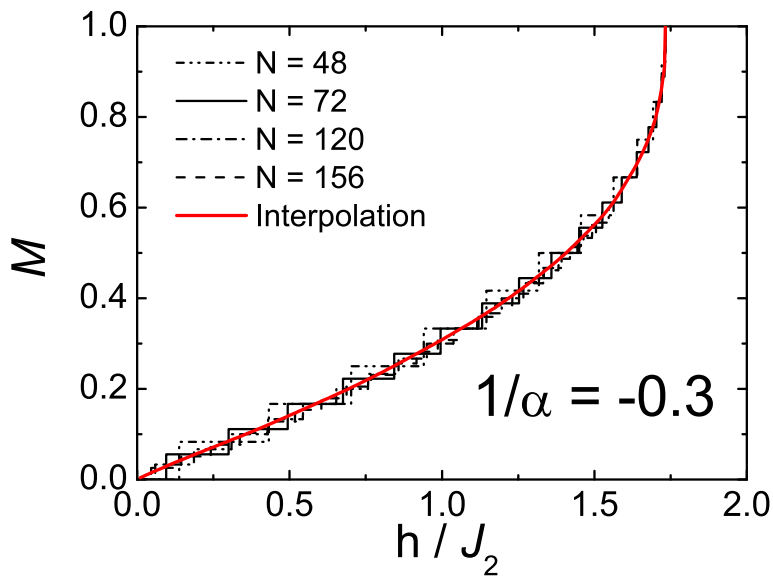}
\caption{DMRG calculations of the magnetisation of a one-
dimensional $S$=$\frac{1}{2}$ Heisenberg chain using $1/\alpha$ =
-0.3 for N = 48, 72, 120 and 156 sites. Solid red line is an
interpolation taking the midpoint of each plateau in the
magnetisation for $N$ = 72 sites.} \label{fig:magcalc}
\end{center}
\end{figure}

Fig~\ref{fig:licuvo4hfc} shows the magnetisation as a function of
applied magnetic field up to 55 T for $H \mid\mid c$ at 1.7K.
Differentiating the magnetisation with respect to the field shows
a saturation of the magnetisation at 40.7 T \cite{ME:exlicuvo4}.
Using the results of the DMRG calculations, $H_{sat}$ = 1.73
$J_{2}$, and the dominant exchange integral, $J_{2}$ $\approx$ 41
K with a $g$-factor of $g_{c}$ = 2.313 \cite{krug:licuvo4} we
arrive at a saturation field of $H^{c}_{sat}$ = 46.2 T. This gives
a $\sim$ 4 T discrepancy in the saturation field between theory
and experiment. The insert of Fig~\ref{fig:licuvo4hfc} shows a
clear anomaly in the derivative of the magnetisation at $\sim$ 7.5
T. To investigate whether this corresponds to a MFCS we have
carried out DMRG calculations in the parameter regime ($1/\alpha$
=$J_{1}$/$J_{2}$ = -0.3) relevant for LiCuVO$_{4}$ as deduced by
INS experiments, based on the following one dimensional
hamiltonian.

\begin{equation}
H = \sum_{i} (J_{1} \vec{S}_{i} \cdot \vec{S}_{i+1} + J_{2}
\vec{S}_{i} \cdot \vec{S}_{i+2}) - h \sum_{i} S_{i}^{z}
\label{eq:hamj1j2}
\end{equation}

Where $\vec{S}_{i}$ are spin 1/2 operators, $J_{i}$, $i$=1,2, are
the exchange integrals and $h$ is the magnetic field.

\begin{figure}
\begin{center}
\includegraphics[bb = 16 13 314 241]{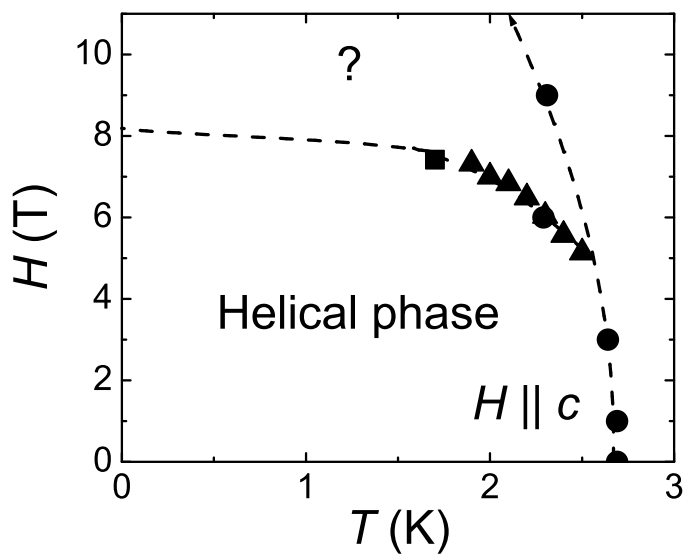}
\caption{Phase diagram of LiCuVO$_{4}$ with $H \mid \mid c$.
Closed circles: specific heat measurements. Closed triangles:
magnetic susceptibility measurements. Closed squares: pulsed high
field measurements. Dashed lines are tentative phase boundaries.}
\label{fig:pd}
\end{center}
\end{figure}


The results of our calculations for 4 different system sizes, $N$
= 48, 72, 120 and 156 are shown in Fig~\ref{fig:magcalc}. In all
cases we see no change of the magnetisation by a jump or
singularity in the middle field region. An interpolation of the
magnetisation for $N$ = 72 using the midpoint of every plateau
\cite{dc:zigzag} results in a smooth curve. Differentiating this
curve with respect to the field (d$M$/d$h$) does not indicate any
singularities or jumps. Any experimental uncertainty in the values
of $J_{1}$ and $J_{2}$ does not affect this result as for 0 $>$
$J_{1}$/$J_{2}$ $\gtrsim$ -1, the frustrated ferromagnetic
$J_{1}$-$J_{2}$ chain exhibits only one phase in a finite magnetic
field until saturation is reached \cite{hm:magchain}. Therefore,
we conclude in the parameter range of LiCuVO$_{4}$ we have no
occurrence of a middle field cusp singularity within the
$J_{1}$-$J_{2}$ model.

Failing the observation of a MFCS in the DMRG calculations, we
suggest the anomaly most likely represents a reorientation of the
magnetic structure. To gain more insight into this we have
constructed  the low temperature phase diagram of LiCuVO$_{4}$ for
$H \mid \mid c$ as deduced by three different experimental
techniques, specific heat, high field magnetisation and magnetic
susceptibility. The data points from the specific heat, as shown
in Fig~\ref{fig:pd}, were taken from the maximum in $C_{p}$ when
passing through the sharp transition temperature at a constant
magnetic field. The data points for the susceptibility were taken
from $M-H$ scans at constant temperature. The differential $dM/dH$
was calculated numerically and the maximum of the peak was taken
as the magnetic field of the transition. The data points for the
high field magnetisation was taken in a similar way to the
magnetic susceptibility. Two distinct phases are shown in
Fig~\ref{fig:pd}, the first is the helical long range order phase
as solved by neutron diffraction. The second phase observed for
$H$ $\geq$ 7.5 T, increases in field with decreasing temperature
and then saturates to reach, at $T$ = 0, approximately 8 T as
shown in Fig~\ref{fig:pd}. To deduce the nature of this phase from
the magnetic structure, a field of $H >$ 7.5 T would be quite
large to represent a spin flop phase resulting from local
anisotropies, therefore the change in the magnetisation in this
phase most likely represents an essential reorientation of the
magnetic structure from the $ab$ plane helix or alternatively a
spin liquid state. Further neutron scattering experiments are
needed in order to investigate the interesting phase diagram of
LiCuVO$_{4}$ in terms of its magnetic structure in applied fields.

\section{Summary}

We have shown that the frustrated one dimensional quantum
antiferromagnet LiCuVO$_{4}$ has a complex ($H-T$) phase diagram.
With the application of a magnetic field larger than $\sim$ 7.5 T
at 1.7 K with $H \parallel c$, induces an anomaly in the
derivative of the magnetisation. Our DMRG calculations, by using a
frustration parameter of 1/$\alpha$ = -0.3 as derived from
inelastic neutron scattering, showed no evidence of a middle field
cusp singularity. The anomaly seen in the derivative of the
magnetisation at $\sim$ 7.5 T could indicate a significant change
of the magnetic structure from the $ab$ plane incommensurate
magnetic helix seen in zero field. Further neutron investigations
are planned in order to elucidate the magnetic structure of this
unknown high field phase.

\ack

F. H.-M. acknowledges support from the NSF grant DMR-0443144. We
thank E. Br\"{u}cher and G. Siegle for experimental assistance.

\section*{References}


\bibliography{LiCuVO4}

\end{document}